\documentclass{PoS}

\title{Measurement of the $\nu_{\mu}$ flux and inclusive charged current cross section at T2K's near detector}

\ShortTitle{$\nu_{\mu}$ flux and inclusive charged current cross section}

\author{\speaker{Melody Ravonel Salzgeber}\thanks{on behalf of the T2K collaboration}\\
        University of Geneva\\
        E-mail: \email{melody.ravonel@cern.ch}}

\abstract{ We present the first measurement of the muon 
neutrino spectrum at the T2K near detector, ND280, using the data collected at the JPARC accelerator facility in Tokai, Japan. ND280 is located 280 meters 
downstream from the target and 2.5$^\circ$ off-axis from the direction of
the beam. The measured spectrum at ND280 constrains the flux and cross section uncertainties in the T2K oscillation analysis. We select inclusive charged-current (CC) events from muon neutrinos in ND280.
These are separated into a charged current quasi-elastic (CCQE) enhanced sample and a CC non-QE sample. We then fit the muon spectrum for both 
samples to extract flux and cross section parameters, which are used as inputs into T2K's oscillation analysis. We separately produce a flux-averaged differential inclusive CC cross-section in the 2-dimensional plane of muon momentum and angle. For the cross-section 
measurement the flux is given by the MC and tuned to data from the
NA61 experiment. We present the event selection, detector uncertainties,
and final measurement result for both the spectrum and for the 
cross-section.}

\FullConference{36th International Conference on High Energy Physics,\\
		July 4-11, 2012\\
		Melbourne, Australia}

\begin{document}

\section{The T2K experiment}
T2K is a neutrino oscillation experiment that is described in detailed in \cite{Abe:2011ks}.
It contains a far detector, Super-Kamiokande (SK), at 295 km from the neutrino source and a near detector at 280 m.  In the near detector complex, there is  an on-axis detector INGRID and a  2.5$^\circ$ off-axis detector, ND280, which is used to measure the neutrino charged curent (CC) rate and cross section in this paper.
Neutrinos are generated from the 30 GeV J-PARC (Japan Proton Accelerator Research Complex) proton beam located at Tokai-mura, Japan. 
The protons are extracted and interact with a graphite target. The positively charged hadrons produced in the interactions (mostly pions and kaons) are focused by three magnetic horns and directed into a 96 m long decay pipe. Here most of the pions and kaons decay, mainly into muons and muon neutrinos. 

ND280 is a multi-purpose set of sub-detectors installed inside a 0.2 T dipole magnet (recycled from the UA1 and NOMAD experiments at CERN). 
 
The tracker is the main component used for the results presented here. It comprises three TPCs and two Fine Grained Detectors (FGDs).
For this measurement, the first FGD has been used as active target for neutrino interactions, while the second FGD has only been used as a tracking detector. 
The composition of the first FGD is mainly Carbon at 86\%  (C$_{86\%}$, O$_{3.7\%}$, H$_{7.4\%}$, Ti$_{1.7\%}$, Si$_{1\%}$, N$_{0.1\%}$), while the second FGD is a water-rich detector.

\begin{figure}[!h]
    \centering
    \includegraphics[width=0.5\textwidth]{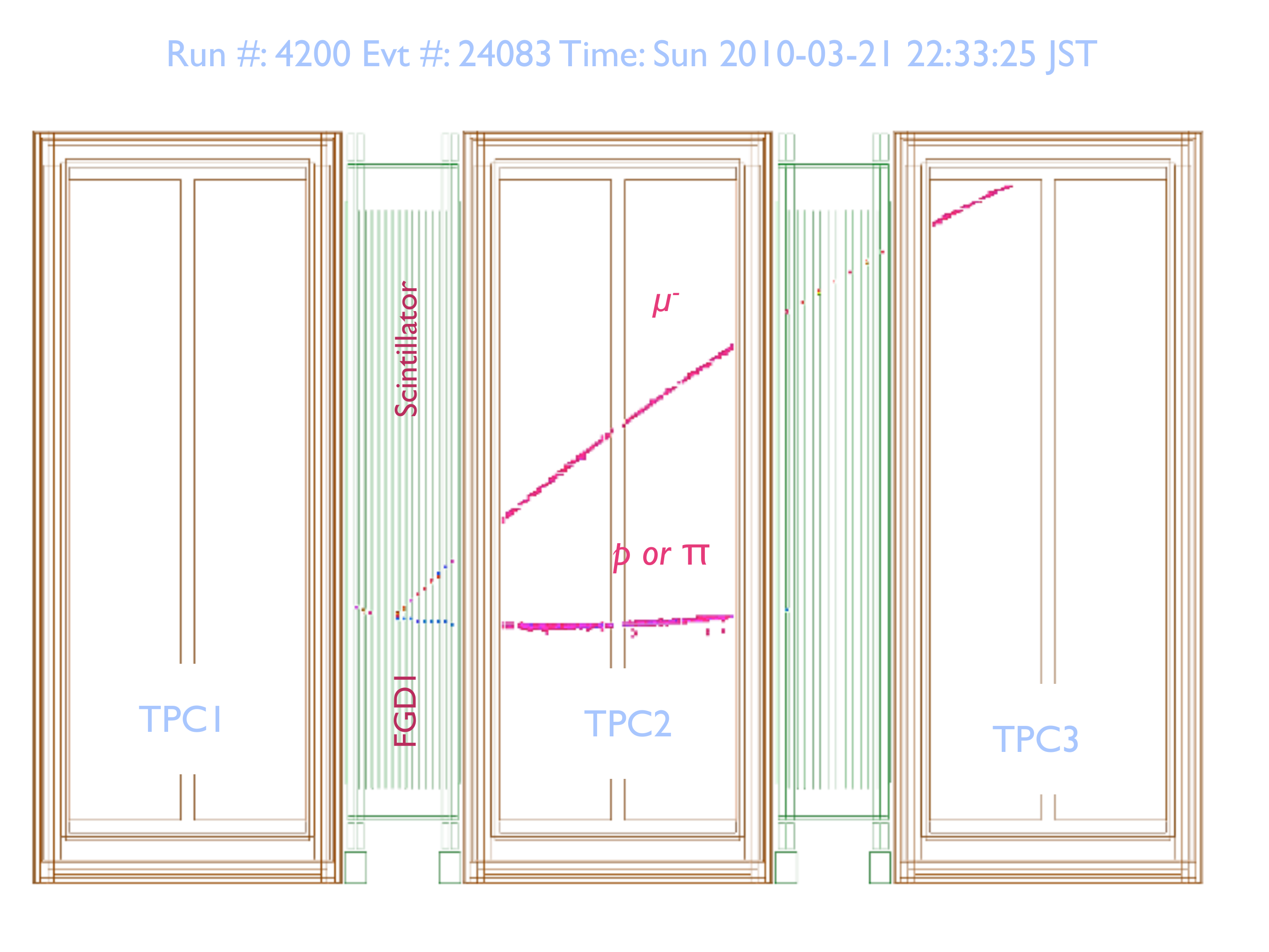}
    \caption{Charged current event candidate in the tracker region of the near detector. }
    \label{fig:ccevent}
  \end{figure}

The TPCs perform three key functions in the near detector.
Firstly, they reconstruct charged particle crossing the detector in three dimensions. Secondly, they measure the momentum of the charged particle thanks to the magnetic field. Thirdly, they provide particle identification using the amount of ionization left by each particle combined with the measured momentum.
These three functions allow the selection of high purity samples of different types of neutrino interactions.  
Fig.~\ref{fig:ccevent} shows a charged current event candidate in the tracker of ND280.

\vspace{-0.5cm}
\section{Event samples, selections and systematic errors}
The same event samples is used to obtain the inclusive charged current (CC) interaction rate and cross section measurement in the tracker. 
From January 2010 up to the major Earthquake in Japan in March 2011, a total of 10.796 $\times$ $10^{19}$ protons on target (POT) have been used.

The inclusive charged current selection is done by first selecting at least one negative track in the TPC. This track should start in the fiducial volume of the first FGD and have an energy loss compatible with the muon hypothesis. 
With this selection a total of 4485 data events have been selected as inclusive charged current interaction candidates.

\begin{figure}[!h]
    \centering
  \includegraphics[width=0.49\textwidth]{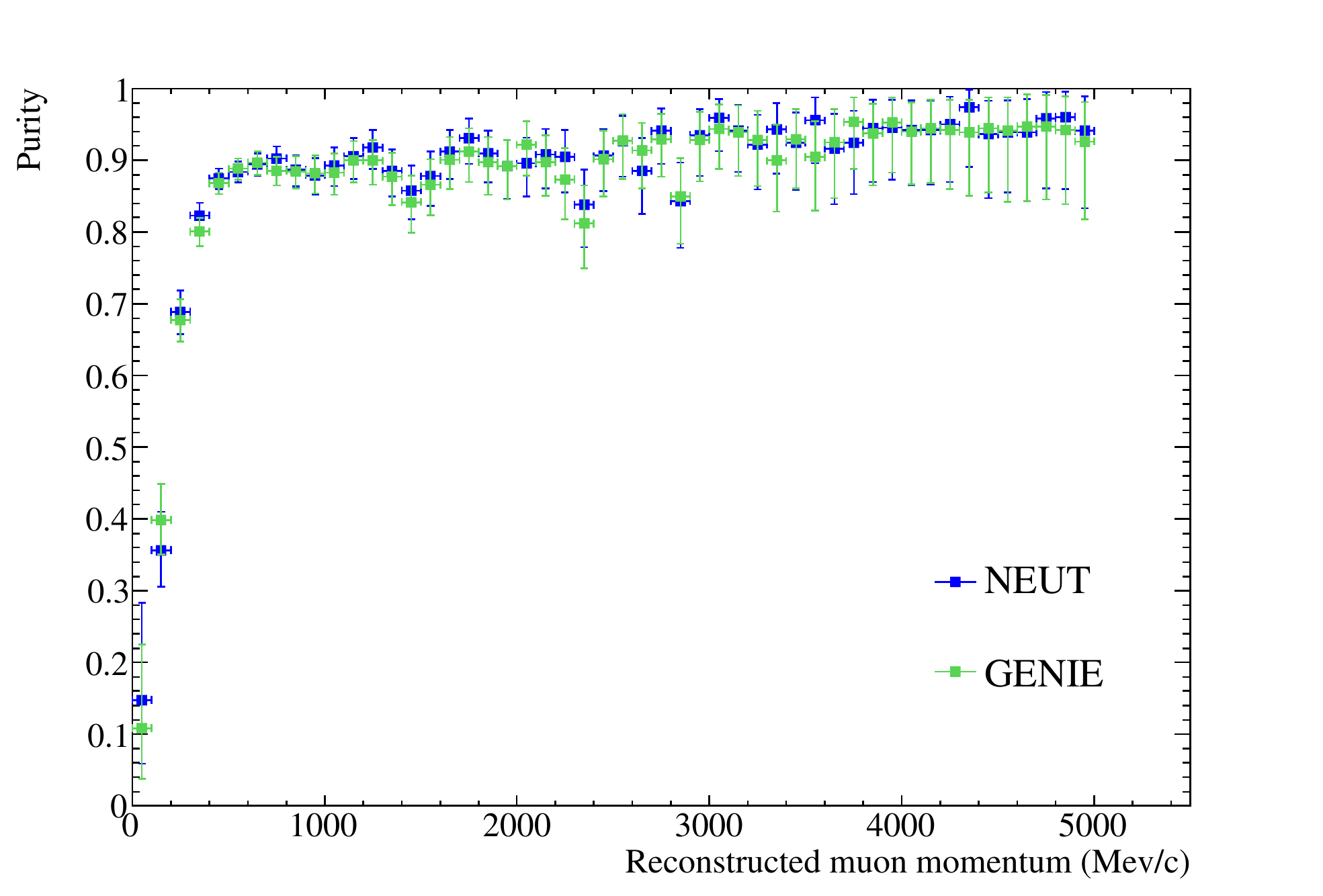}	
  \includegraphics[width=0.49\textwidth]{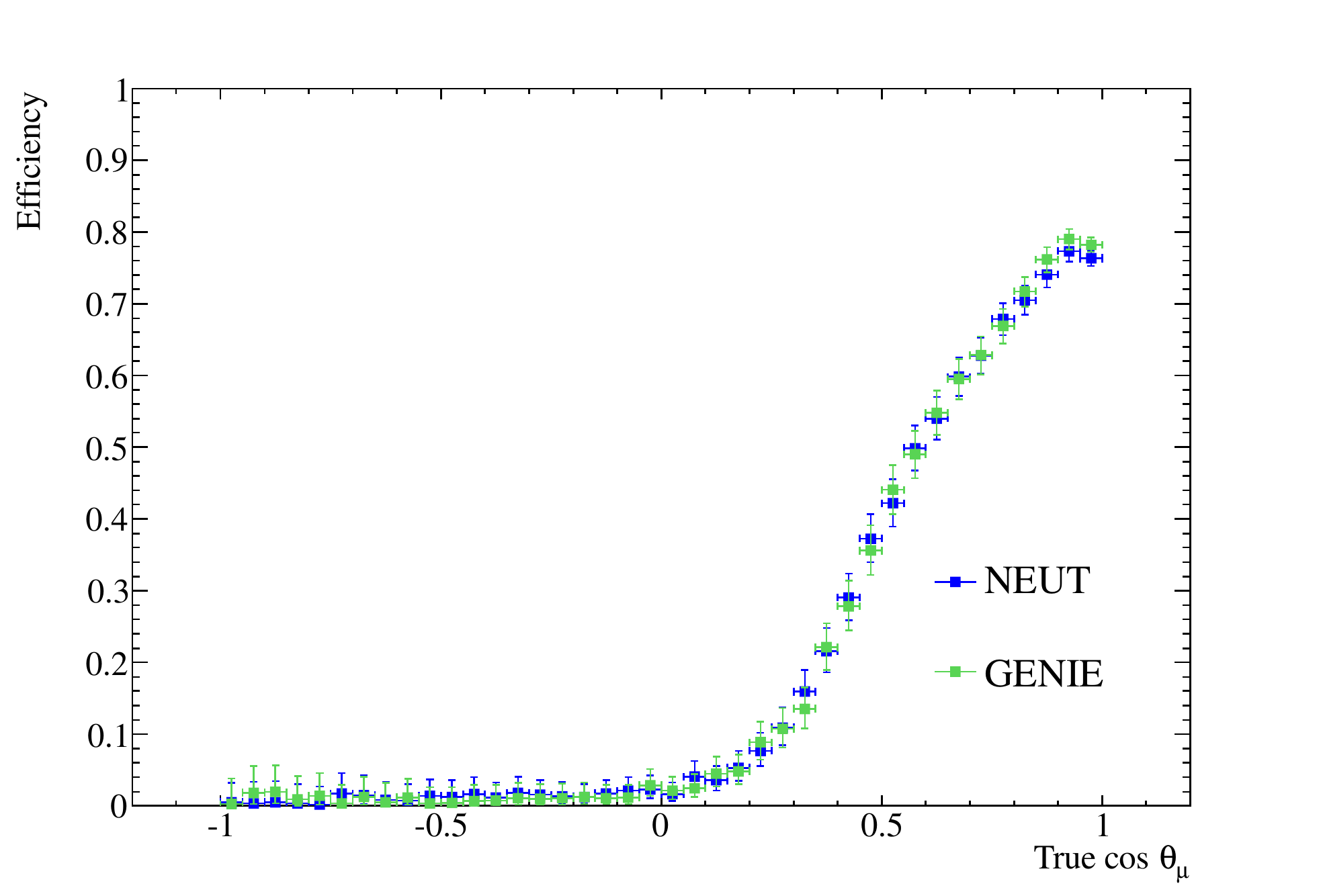}
        \caption{Purity vs the muon momentum (left) and efficiency vs muon angle (right).}
    \label{fig:eff}
  \end{figure}

 An additional step to this basic selection is then done to distinguish the quasi-elastic (CCQE) interactions in the case of the interaction rate measurement. The quasi-elastic sub-sample is obtained by requiring in addition the existence of only one TPC-FGD track and the absence of any Michel electron.
Comparison with Monte-Carlo (MC) simulations has been done using two neutrino interaction generators, NEUT \cite{Hayato:2009zz} and GENIE \cite{Andreopoulos:2009zz}.
The final purity and efficiency are 72\% and 40\%, respectively, for the CCQE sample and 87\% and 50 \% for the CC sample.
As can be seen in Fig.~\ref{fig:eff}, this selection is optimized for forward-going muons, and the purity is 
very high starting from  $\sim$ 500 MeV.
  
Most of the systematic error sources are the same for the rate and the cross section measurements. They can be subdivided in three independent categories: flux prediction uncertainties, cross section model uncertainties and detector response uncertainties. 
For the cross section measurement there are two additional systematic errors, namely the number of target uncertainty in the first FGD and the unfolding uncertainty.
The flux uncertainty is the main systematic error due to the big uncertainty in hadron production cross sections and secondary interactions. The hadron production in proton-Carbon interactions at 30 GeV has been measured by the NA61/SHINE collaboration \cite{Abgrall:2011ae} and used for neutrino flux prediction in T2K.

The cross section model uncertainties are estimated comparing our main generator NEUT with the MiniBOONE data. The MiniBooNE data corresponds to the same range of energy as T2K and presents no angular dependence due to its spherical design. 
The main contribution to the detector systematic uncertainty is coming from the out of fiducial volume background and momentum distortions due to the magnetic field.
The systematic error due to the algorithm is less than 1\% for all forward bins, while the number of target uncertainty represents an overall uncertainty of 6.7\%. 

\vspace{-0.2cm}
\section{Muon neutrino interaction rate measurement}
The rate measurement method is based on a Likelihood fit on the tracker charged current data. The number of reconstructed events at the near detector is fitted for 40 angular and momentum bins, where the first 20 bins are for the CCQE selection, and the last 20 bins are for the CCnQE. 

 The likelihood is constructed from the binned number of events in data and prediction, $N_i^{data}$ and $N_i^{MC}$, where $i$ is the one-dimensional bin index.% corresponding to the 4 (angular) $\times$5 (momentum) bins.
The likelihood depends on parameters that describe the uncertainties in the underlying flux and cross section models, $\vec{b}$ and $\vec{x}$. The likelihood also depends on nuisance parameters that model the detector systematic uncertainties, $\vec{d}$.

The quantity that is maximized is the likelihood ratio,
\begin{eqnarray}
\mathcal{L}_{ND280}=\frac{\pi(\vec{b})\pi(\vec{x}) \pi(\vec{d}) \displaystyle\prod_i^{p-\theta~bins} N^{MC}_i(\vec{b},\vec{x},\vec{d})^{N_i^{data}}e^{-N_i^{MC}(\vec{b},\vec{x},\vec{d})}/N_i^{data}!}{\pi(\vec{b}_{nom})\pi(\vec{x}_{nom}) \pi(\vec{d}_{nom})\displaystyle\prod_i^{p-\theta~bins} (N_i^{data})^{N_i^{data}}e^{-N_i^{data}}/N_i^{data}!}
\end{eqnarray}

Systematic errors are taken into account inside the fit by the use of a prior probability density function (PDF) defined similarly for detector, cross section and flux parameters,
\begin{eqnarray}
\pi(s)=\frac{1}{(2\pi)^{k/2}|V_s|^{1/2}}e^{(-\frac{1}{2}\Delta \vec{s}{V_s^{-1}}\Delta\vec{s}^{~T})}
\end{eqnarray}
where $\Delta\vec{s}$ corresponds to the set of fitted parameters that is related to a given source of systematic uncertainty, $V_s$ is the covariance matrix for the given systematic uncertainty, and $k$ the dimension of the parameter vector $\vec{s}$.

\begin{figure}[!h]
    \centering
    \includegraphics[width=0.49\textwidth]{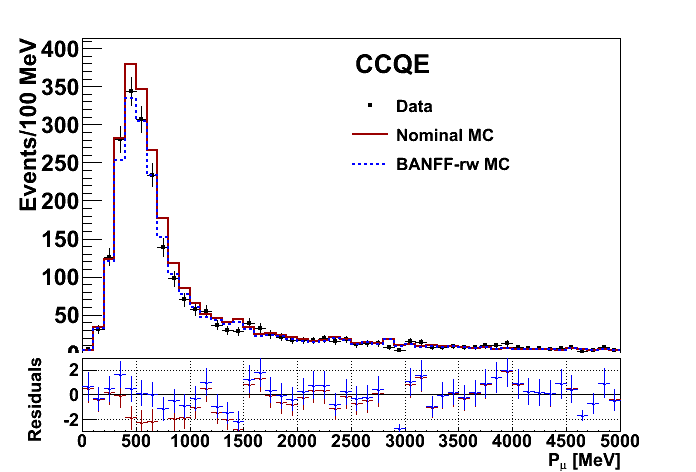} 
     \includegraphics[width=0.49\textwidth]{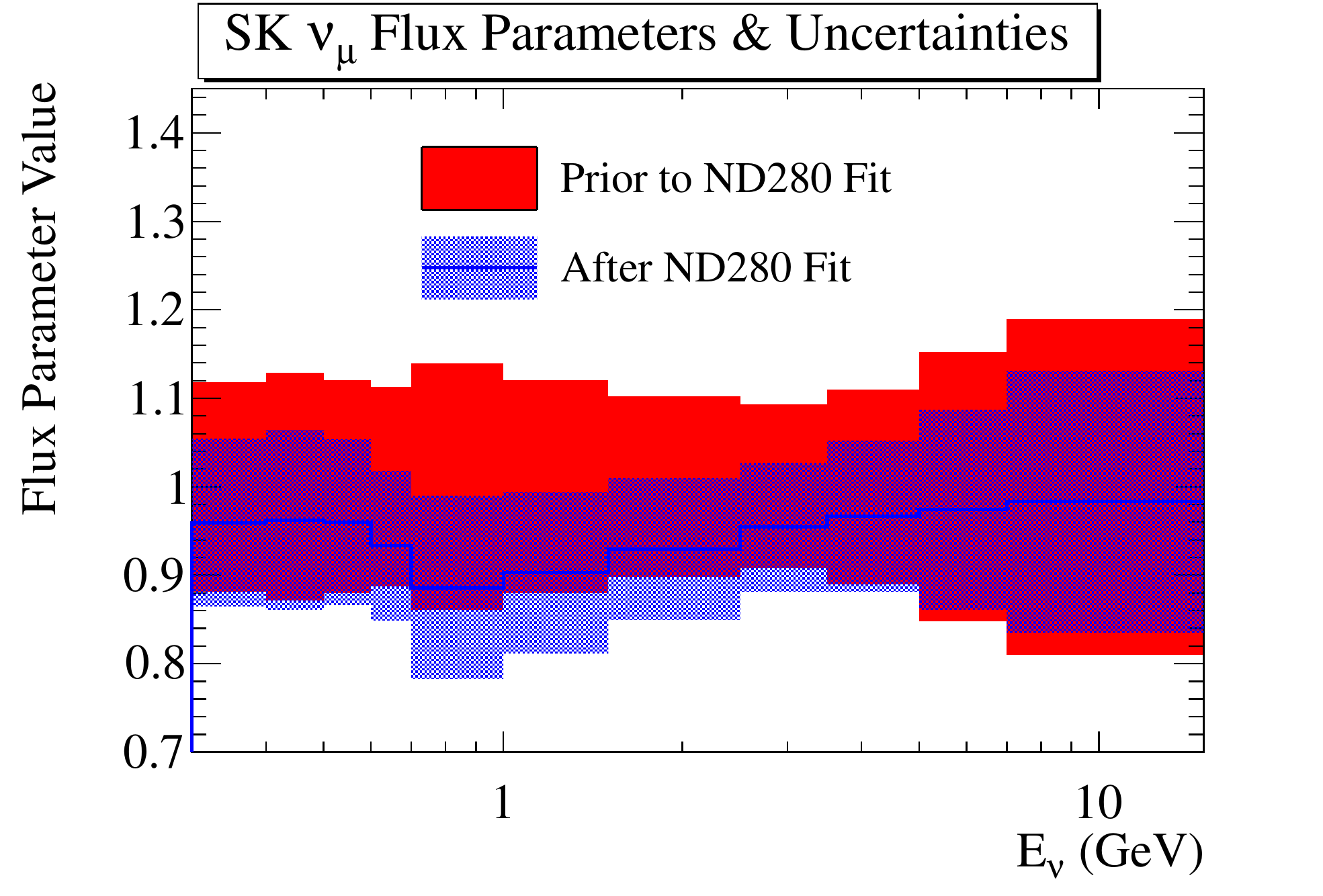}
        \caption{Comparison between nominal and reweighted MC for the muon momentum. The post-fit result is shown in blue and labeled as BANFF-rw MC (left). The SK $\nu_{\mu}$ flux parameters with 1$\sigma$ uncertainties
before and after the near detector $\nu_{\mu}$ fit. The flux 
 }
    \label{fig:SKresult}
  \end{figure}

The result of the fit gives a much better agreement between data and MC. While the data to MC ratio is 95.0 \% and 98.7\% for the CCQE selection and CCnQE selection for the nominal MC, the ratio becomes 99.9 \% and 99.4 \% after fit as can be seen in Fig.~\ref{fig:SKresult}.

In addition to a better data to MC agreement after fitting, the fit to the near detector data 
constrains the flux and the cross section at the far detector.
The systematic error of the neutrino flux parameters at SK is reduced with the use of the ND280 data fit. In particular, the total systematic error on the flux parameters is decreased from 20\% before the ND280 fit to 10\% after the ND280 fit. 
  
\vspace{-0.2cm}
\section{Muon neutrino cross section measurements}

Based on Bayes' theorem, the method employed to extract the cross section is a 2-Dimensional unfolding of the muon momentum and angle reconstructed distributions.
The result of the unfolding gives the number of inferred events in ``true'' bins, where we correct the smearing due to the response of the detector.  
The binning used for the reconstructed distributions is the same as for the rate measurement.
As there are very few tracks reconstructed as going backward, the first reconstructed bin in angle takes into account the entire backward region up to $\cos \theta_{\mu}=0.84$. To account for the fact that backward going tracks are badly reconstructed, the binning on the inferred number of events is slightly different and splits the first angular bin into two. The two bins are the backward region and the forward region up to $\cos \theta_{\mu}=0.84$.

\begin{figure}[!h]
  \centering
  \includegraphics[width=0.49\textwidth]{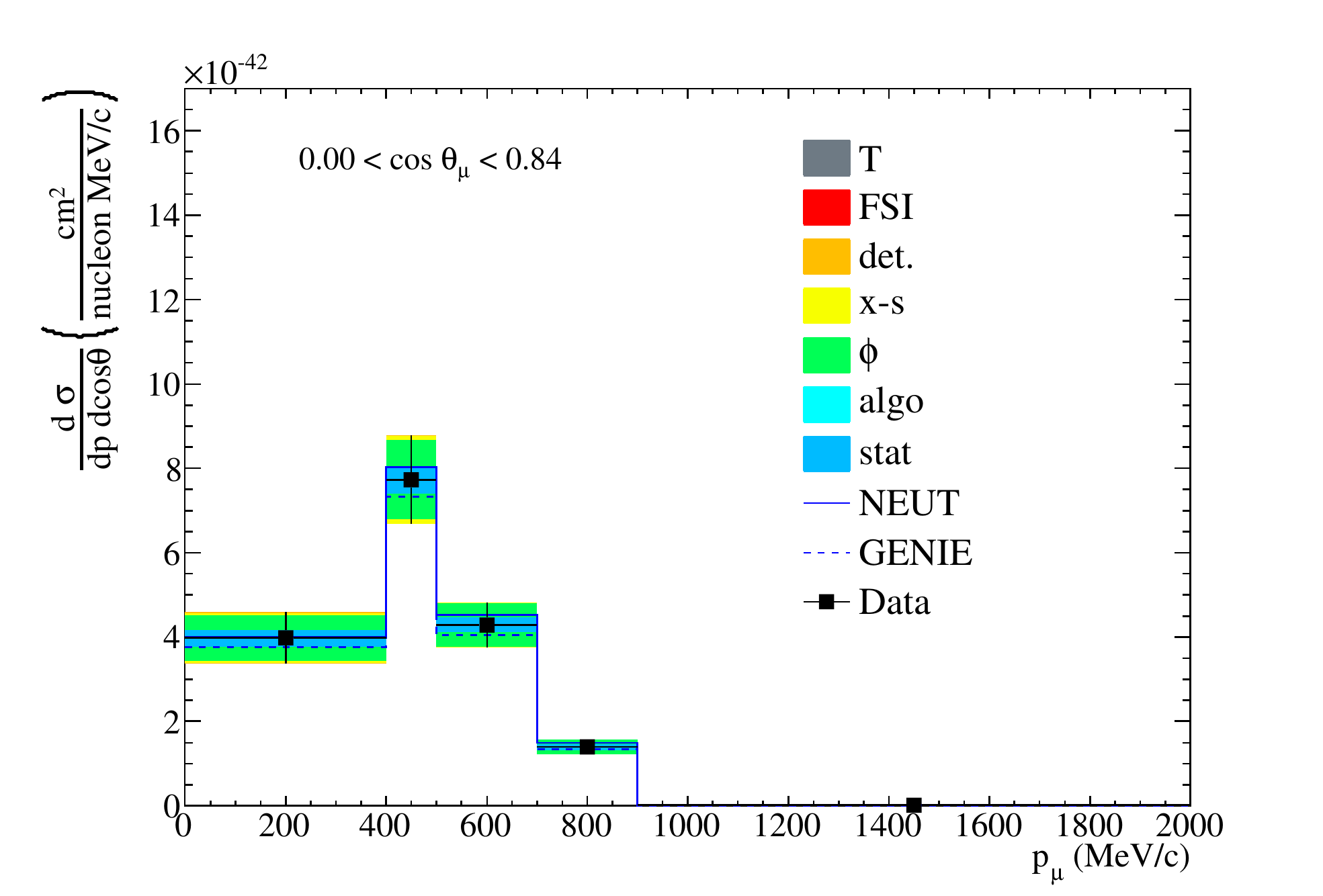}
  \includegraphics[width=0.49\textwidth]{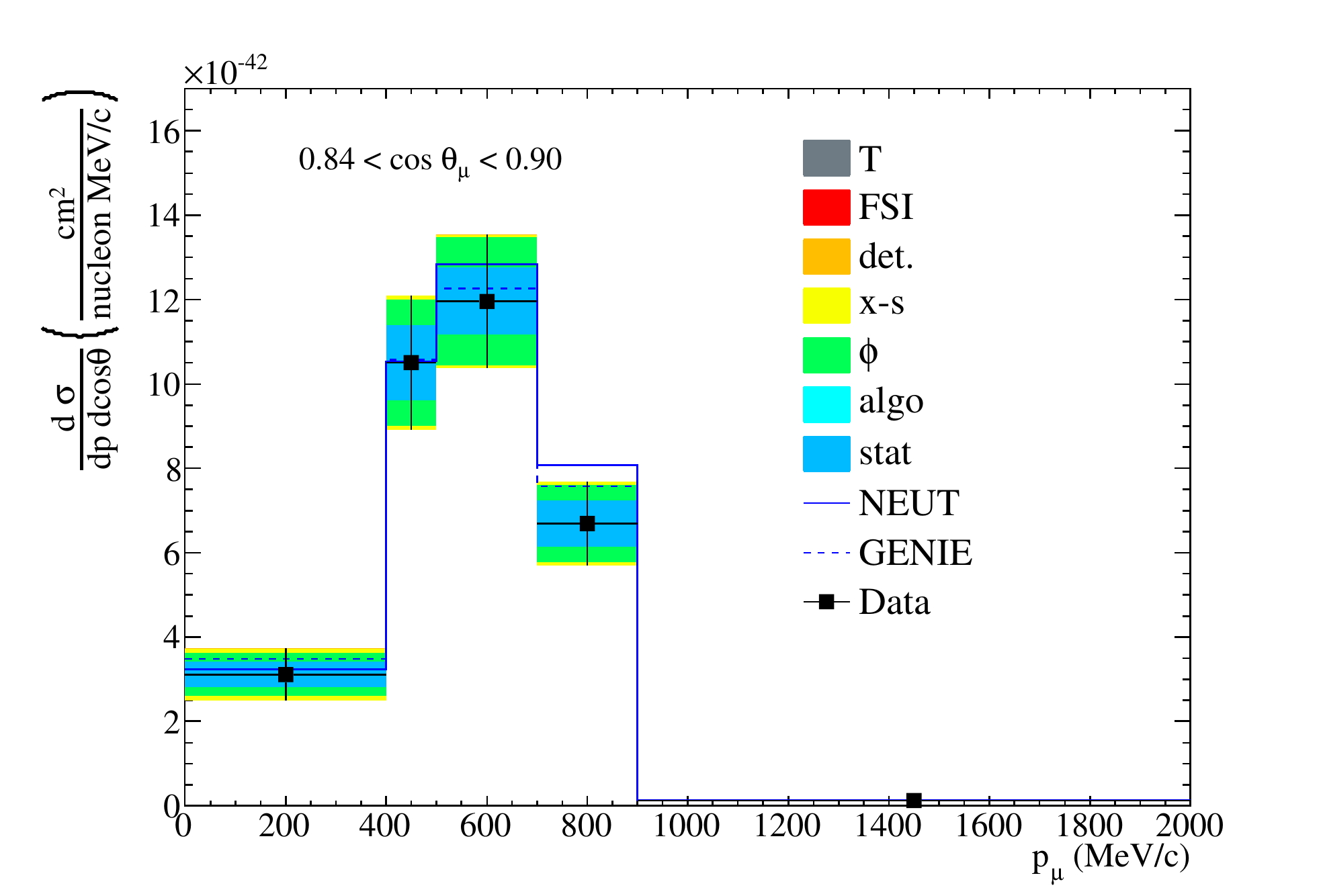}
  \includegraphics[width=0.49\textwidth]{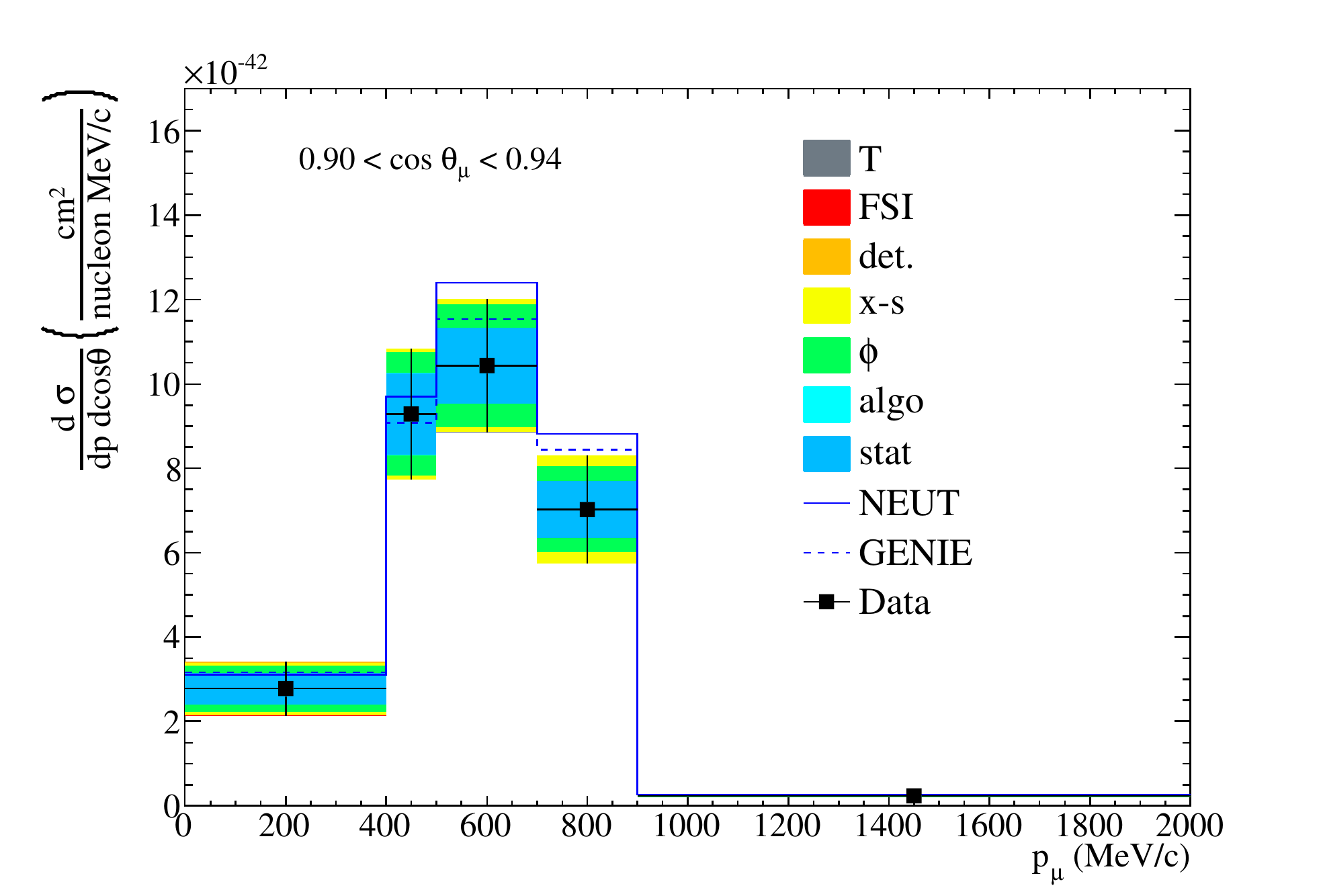}
  \includegraphics[width=0.49\textwidth]{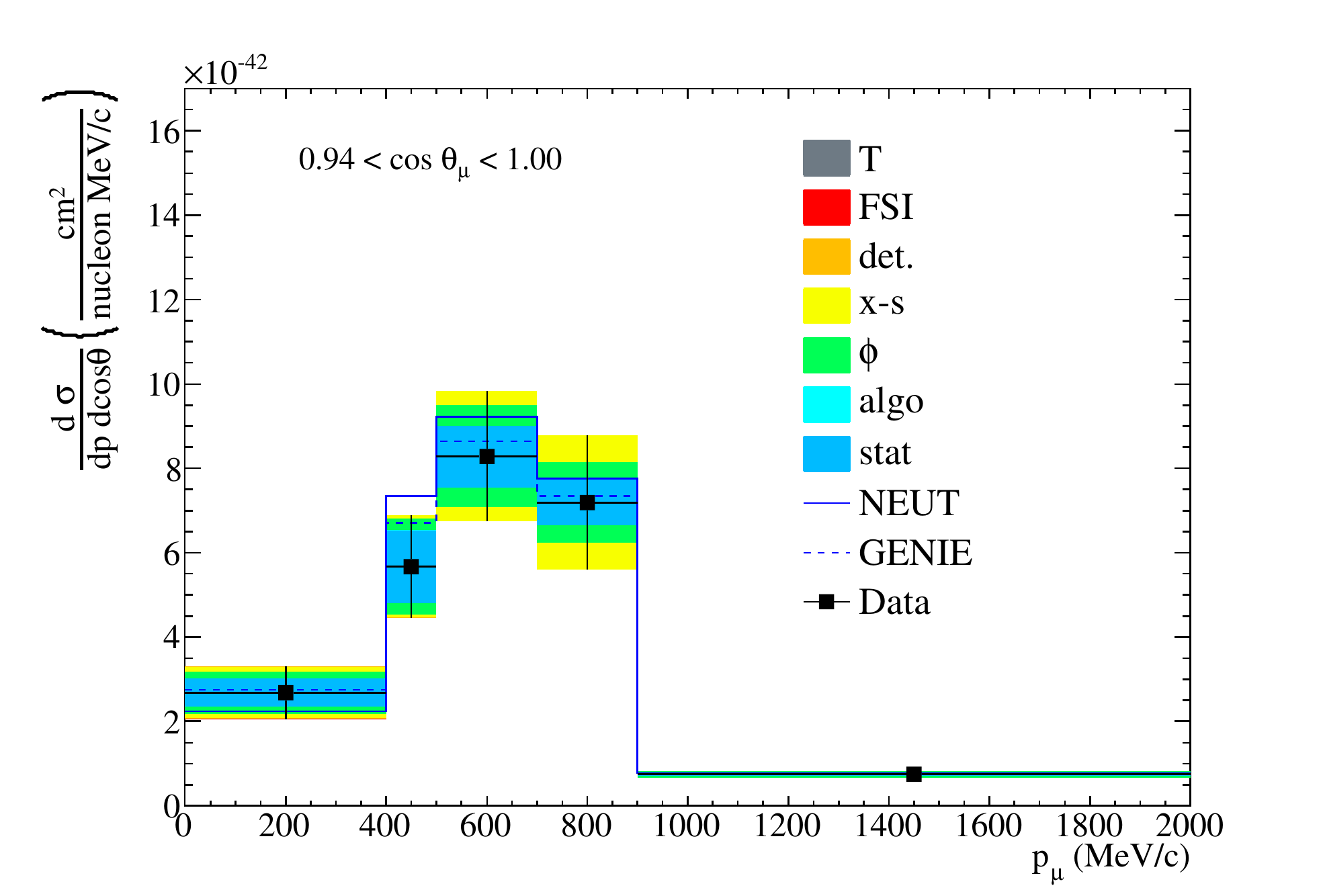}
  \vspace{-0.3cm}
  \caption{The differential cross section results with systematic and statistical error bars together given in cm$^2$/nucleon/MeV. Each graph corresponds to a bin angle. Each color represents a systematic error source, T: number of Target uncertainty, FSI: Final State Interaction uncertainty, det.: detector uncertainty, x-s: cross section modeling uncertainty, $\phi$: flux uncertainty, algo: algorithm uncertainty, stat: statistical error. }
\label{fig:dxs}
\end{figure}

The final cross section result is a flux-averaged double differential cross section in muon momentum and angle. It is obtained from the result of the unfolding divided by the total number of target nucleons, the total flux of the incoming muon neutrinos and by the bin widths. The final result is shown in  Fig. \ref{fig:dxs}.
\vspace{-0.3cm}

\begin{figure}[tbh]
  \centering
  \includegraphics[width=0.65\textwidth]{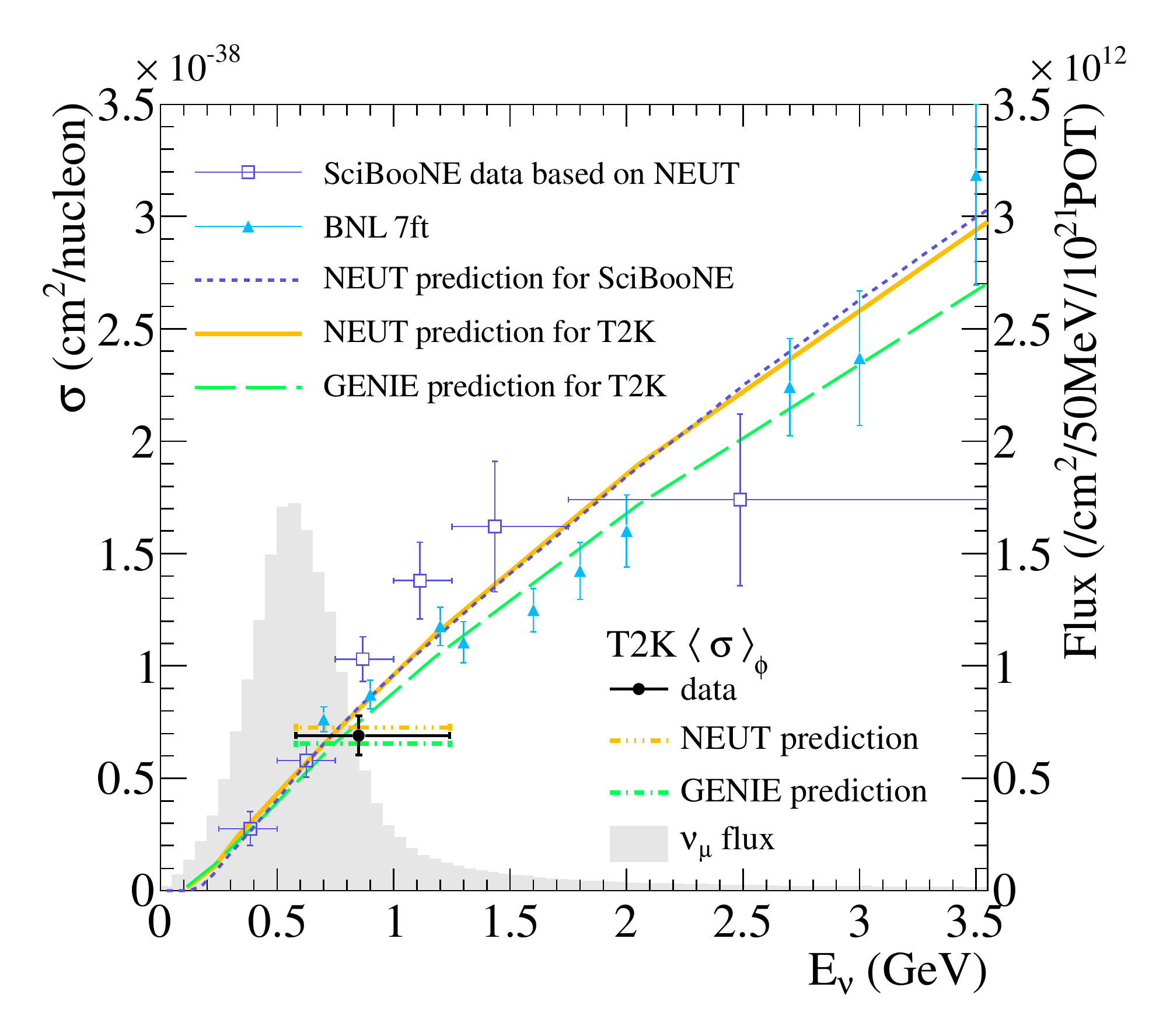}
  \caption{The T2K total flux-averaged cross section with the NEUT and the GENIE prediction for T2K and SciBooNE. The T2K data point is placed at the flux mean energy. The vertical error represents the total (statistical and systematic) uncertainty, and the horizontal bar represent 68\% of the flux at each side of the mean energy.The T2K flux distribution is shown in grey. The predictions for SciBooNE have been done for a C$_8$H$_8$ target \cite{Nakajima:2010fp} which is comparable to the mixed T2K target. BNL data has been measured on deuterium \cite{Mukhin:1979bd}.}
\label{fig:res}
\end{figure}

Systematic and statistical errors have been propagated through the algorithm by reweighting the MC for each source of error separately. 
The main systematic error source is the muon neutrino flux. In particular the hadron production cross section at the target and the multiplicity of kaons and pions that decay into muon neutrinos.
The cross section model uncertainty is also very important. In particular, we changed the nucleon distribution inside the nucleus in the FGD1, from the simple Relativistic Fermi Gas Model to the more sophisticated Spectral Function of the NuWro generator \cite{Golan:2012wx}. The difference of the two results has been taken as systematic error and mainly affect forward-going muons with relatively high momentum. The other systematic uncertainties are almost negligible in comparison.
Fig.~\ref{fig:res} shows the final double differential cross section only for forward-going muons, since the efficiency of selecting charged current interactions with backward-going muons is very small.
\vspace{-0.3cm}

 A flux averaged total cross section is also provided by extrapolating to the backward going bin. The final result is shown in Fig. \ref{fig:res}. 
The total flux averaged cross section result for our target is: 
$\langle \sigma_{ \rm{CC}}^{\rm{DATA}} \rangle_{\phi} =(6.91 \pm 0.13 (stat) \pm 0.84 (syst)) \times 10^{-39} \rm{\frac{cm^2}{nucleon}}$

\vspace{-0.2cm}
\section{Conclusion}
The T2K collaboration has presented a notable improvement in the precision of the predicted event rates at the T2K far detector based on the measurements at the near detector. A reduction of the systematic error on the flux parameters at the far detector from 20\% to 10\% has been estimated due to the new near detector data. 
In addition to constraining flux and cross section models at the far detector, the near detector data has also been used to extract the charged current inclusive differential cross section.
While the results presented here are all preliminary, publications on the various topics are in preparation.

\vspace{-0.2cm}
\bibliography{skeleton}

\end{document}